\documentstyle[osa,10pt]{revtex}

\begin{document}

\title{Molecular photon echoes as a signature of vibrational cat states}

\author{Holger F. Hofmann, Takao Fuji, Atsushi Sugita, and Takayoshi Kobayashi
\\Department of Physics, Faculty of Science, University of Tokyo,\\
7-3-1 Hongo, Bunkyo-ku, Tokyo113-0033, Japan}

\maketitle

\begin{abstract}
Quantum interference between two distinct vibrational trajectories 
induced by two pulse femtosecond excitation in molecules is shown to
result in a photon echo, providing direct evidence
of the cat state superposition of Gaussian vibrational wavepackets.
\\
(OCIS codes: Atomic and molecular physics;(270.0270) Quantum Optics;
(020.1670) Coherent optical effects)
\end{abstract}
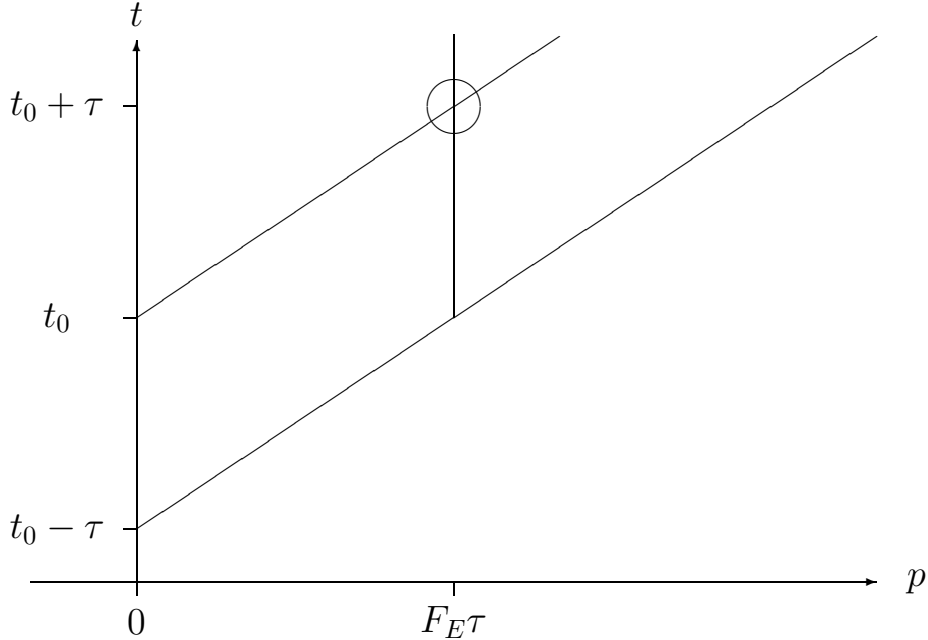
\begin{figure}[hb]
\begin{picture}(470,270)
\put(70,40){\vector(1,0){320}}
\put(395,30){\makebox(20,20){\Large $p$}}
\put(230,35){\line(0,1){5}}
\put(220,15){\makebox(20,20){\Large $F_E\tau$}}
\put(100,15){\makebox(20,20){\Large 0}}
\put(110,35){\vector(0,1){210}}
\put(100,245){\makebox(20,20){\Large $t$}}
\put(105,60){\line(1,0){5}}
\put(60,50){\makebox(40,20){\Large $t_0-\tau$}}
\put(105,140){\line(1,0){5}}
\put(60,130){\makebox(40,20){\Large $t_0$}}
\put(105,220){\line(1,0){5}}
\put(60,210){\makebox(40,20){\Large $t_0+\tau$}}
\put(110,60){\line(3,2){280}}
\put(110,140){\line(3,2){160}}
\put(230,140){\line(0,1){107}}
\put(230,220){\circle{20}}
\end{picture}
\caption{\label{path} 
Time dependence of the vibrational momentum $p$ for femtosecond
excitations at $t_0-\tau$ and $t_0$. The excited state is accelerated by a
force $F_E$ equal to the slope of the excited state potential. The circle 
marks the photon echo interference between the ground state and the excited 
state component at $t_0+\tau$.}
\vspace{0.5cm}
\end{figure}

\vspace{0.5cm}
Optical femtosecond pulses can be used to control the quantum coherence 
of the vibrational states in molecules\cite{Koh95,Kis96}. 
In particular, a pair of pulses 
can create the superposition of two distinct coherent states known as
cat states \cite{Jan94}. However, it is usually rather
difficult to obtain experimental evidence of nonclassical quantum 
interference effects from vibrational cat states. 
In order to overcome this problem, it is useful to study the full 
nonlinear excitation dynamics in an approximate model suitable to the short
time dynamics of the molecule. In this presentation, it is shown that
the formation of a vibrational cat state in a molecular transition 
by a pair of femtosecond pulses at times $t_0-\tau$ and $t_0$ gives 
rise to a photon echo at $t_0+\tau$. This photon echo is a signature
of the quantum interference between two distinct vibrational trajectories
in momentum space as shown in figure \ref{path}.

The total vibrational dynamics of a molecule can be described by a 
Hamiltonian $H_0$ which reads
\begin{equation}
\hat{H}_0 = \frac{\hat{p}^2}{2 m} + 
  V_G(\hat{x})\otimes \!\mid\! G \rangle \langle G \!\mid\!
+ V_E(\hat{x})\otimes \!\mid\! E \rangle \langle E \!\mid\!,
\end{equation}
where the vibrational coordinates are represented by the position operator
$\hat{x}$ and the momentum operator $\hat{p}$, and the electronic 
system is represented by a ground state $\!\mid\! G \rangle$ and an excited
state $\!\mid\! E \rangle$. The projection operators 
$\!\mid\! G \rangle \langle G \!\mid\!$ and $\!\mid\! E \rangle \langle E \!\mid\!$
select the corresponding vibrational potentials $V_G(\hat{x})$ 
and $V_E(\hat{x})$. $m$ is the effective mass of the vibration.
Initially, the molecule is in the electronic and vibrational
ground state, centered around a position $x=0$ at the minimum of 
$V_G(\hat{x})$. If the time delay between the two pulses 
creating the cat state is much shorter than the period of the molecular 
vibration, it is possible to neglect the vibrational dynamics,
approximating the potentials by $V_G(0)=0$ and $V_E(0)-F_E(0)\hat{x}$. 
The effective Hamiltonian then reads
\begin{equation}
\label{eq:Ham}
\hat{H}_0 \approx \left(V_E(0) - F_E(0)\hat{x}\right)
\otimes \!\mid\! E \rangle \langle E \!\mid\!.
\end{equation}
This Hamiltonian describes the linear acceleration of the molecular
vibration upon excitation. The temporal evolution of an arbitrary initial 
state may then be given in momentum space by 
\begin{eqnarray}
\psi_G(p;t) &=& \psi_G(p;0)
\nonumber \\
\psi_E(p;t) &=& \psi_E(p\!-\!F_E(0)\,t;\; 0),
\end{eqnarray}
where $\psi_{E/G}(p;t) = \langle E/G ; p \!\mid\! \psi(t)\rangle$ is the
probability amplitude component of the momentum and excitation eigenstate 
$\!\mid\! E/G ; p \rangle $ of the time dependent quantum state 
$\!\mid\! \psi(t)\rangle$. 

Using this simple evolution to determine the molecular dynamics
between a pair of femtosecond pulses, it is a straightforward matter
to describe the creation and evolution of a cat state. Initially,
the molecular ground state is given by $\psi_G(p;0)=\psi_0(p)$ and
$\psi_E(p;0)=0$. At time $t_0-\tau$, the molecule is excited to
$\psi_G(p;t_0-\tau)=\cos(\phi/2)\psi_0(p)$ and
$\psi_E(p;t_0-\tau)=\sin(\phi/2)\psi_0(p)$. The momentum distribution
of the excited state now begins to shift at a rate of $F_E(0)$. 
At $t_0$, a second pulse identical to the first one excites 
the ground state components and
de-excites the excited state components. For $t>t_0$, the 
state of the molecule is given by
\begin{eqnarray}
\psi_G(p;t) &=& \cos^2(\phi/2) \psi_0(p) 
              + \sin^2(\phi/2) \psi_0\left(p\!-\!F\tau\right)
\nonumber \\
\psi_E(p;t) &=& 
\sin(\phi/2)\cos(\phi/2)\left(\psi_0\left(p\!-\!F(t\!-\!t_0)\right)
+ \psi_0\left(p\!-\!F\tau\!-\!F(t\!-\!t_0)\right)\right).
\end{eqnarray}
\begin{figure}[ht]
\begin{picture}(470,200)
\put(35,20){\vector(1,0){400}}
\put(100,15){\line(0,1){5}}
\put(90,0){\makebox(20,10){$0$}}
\put(230,15){\line(0,1){5}}
\put(220,0){\makebox(20,10){$F_E\tau$}}
\put(360,15){\line(0,1){5}}
\put(350,0){\makebox(20,10){$2 F_E\tau$}}
\put(440,15){\makebox(20,10){$p$}}
\put(30,60){\makebox(20,10){$\psi_G(p)$}}
\put(35,140){\vector(1,0){400}}
\put(100,135){\line(0,1){5}}
\put(90,120){\makebox(20,10){$0$}}
\put(230,135){\line(0,1){5}}
\put(220,120){\makebox(20,10){$F_E\tau$}}
\put(360,135){\line(0,1){5}}
\put(350,120){\makebox(20,10){$2 F_E\tau$}}
\put(440,135){\makebox(20,10){$p$}}
\put(30,180){\makebox(20,10){$\psi_E(p)$}}

\put(230,55){\line(0,1){50}}
\put(230,55){\line(1,1){10}}
\put(230,55){\line(-1,1){10}}
\put(230,105){\line(1,-1){10}}
\put(230,105){\line(-1,-1){10}}
\put(250,75){\makebox(20,10){\Large Echo}}

\bezier{200}(60,20)(70,20)(80,60)
\bezier{200}(80,60)(90,100)(100,100)
\bezier{200}(140,20)(130,20)(120,60)
\bezier{200}(120,60)(110,100)(100,100)

\bezier{200}(190,20)(200,20)(210,30)
\bezier{200}(210,30)(220,40)(230,40)
\bezier{200}(270,20)(260,20)(250,30)
\bezier{200}(250,30)(240,40)(230,40)

\bezier{200}(190,140)(200,140)(210,157)
\bezier{200}(210,157)(220,174)(230,174)
\bezier{200}(270,140)(260,140)(250,157)
\bezier{200}(250,157)(240,174)(230,174)

\bezier{200}(320,140)(330,140)(340,157)
\bezier{200}(340,157)(350,174)(360,174)
\bezier{200}(400,140)(390,140)(380,157)
\bezier{200}(380,157)(370,174)(360,174)
\end{picture}
\caption{\label{cat} Schematic illustration of the interference 
between the vibrational cat states associated with the electronic 
excited state and the electronic ground state, respectively.
The vibrational wavepackets are shown in momentum space at $t_0+2\tau$,
following a pair of pulses with $\phi=\pi/3$ at $t_0$ and $t_0+\tau$.} 
\vspace{0.5cm}
\end{figure}
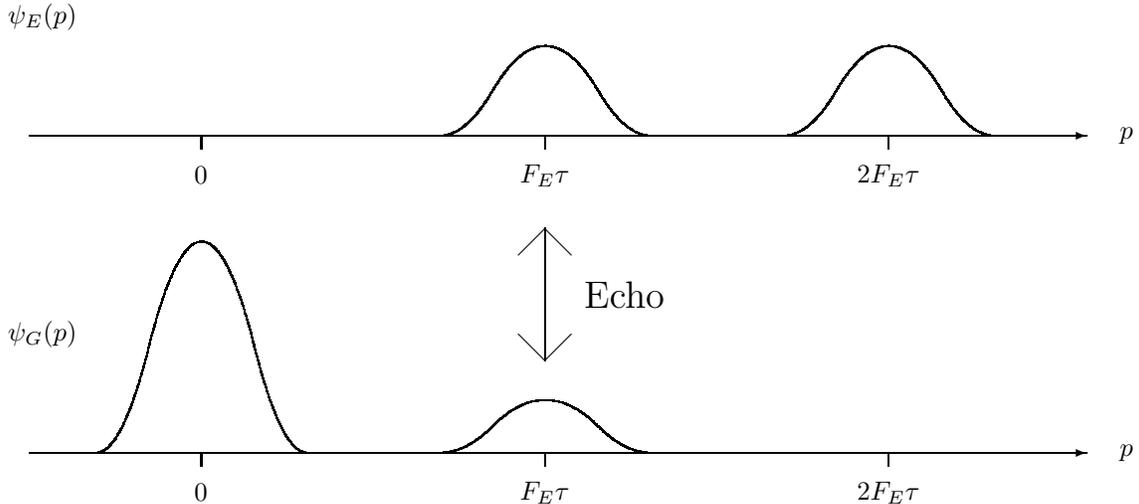
At $t=t_0+\tau$, the momentum shift of the last term in $\psi_G(p;t)$
coincides with the momentum shift of the first term in $\psi_E(p;t)$,
as shown in figure \ref{cat}. The quantum coherence between the excited
and ground state contributions therefore gives rise to a nonvanishing
dipole matrix element. This coherent dipole can be observed as a 
photon echo. 

While the analysis of the quantum dynamics in momentum space clearly 
reveals that the cause of the photon echo is a cat state coherence 
between different velocities of the oscillating atoms, the analogy with
photon echoes in inhomogeneously broadened spectra suggests an alternative
interpretation. Since the approximate Hamiltonian given by equation 
(\ref{eq:Ham}) depends only on the vibrational position $\hat{x}$ and
not on the momentum $\hat{p}$, it is possible to interpret the vibrational 
quantum state as a probability distribution over vibrational position.
The photon echo would then correspond to a rephasing of dipoles associated
with well defined atomic positions. This interpretation is entirely 
sufficient as long as the positional shift induced by the momentum 
differences is indeed negligible. However, the momentum differences become
observable at longer timescales, causing a gradual disappearance of the
photon echo. An analysis of the phase space trajectories suggests a 
decay law of the photon echo intensity $I$ as a function of delay time 
$\tau$ given by
\begin{equation}
\frac{I}{I_0} = \exp\left(- \frac{F^2 \Omega}{2\hbar m} \tau^4\right),
\end{equation}
where $\Omega$ is the angular frequency of the molecular vibration.
Note that the rather unusual $\tau^4$ dependence should allow a clear
distinction between the cat state photon echo in vibrational transitions
and the conventional photon echoes in inhomogeneous systems.

In order to satisfy the requirements for the observation of a cat state
photon echo, a low frequency of the molecular vibration and a strong
electron phonon coupling are necessary. Preliminary studies suggest
that an experimental realization of cat state photon echoes is
possible using the vibrations of the iodine ion in iodine bridged 
metal complexes. In this type of molecular system, the approximations 
applied are valid for time delays $\tau$ of about twenty femtoseconds.
By applying an experimental setup for photon echo spectroscopy using 
femtosecond pulses to this molecular system, it should thus be possible 
to demonstrate cat state interference in molecular vibrations. 
\\[0.5cm]
One of us (HFH) would like to acknowledge support from the Japanese
Society for the Promotion of Science, JSPS.


\begin{thebibliography}{xxx10}
\bibitem{Koh95} B. Kohler et. al., ``Quantum control of Wave Packet
Evolution with Tailored Femtosecond Pulses,'' Phys.Rev.Lett. 
{\bf 74}, 3360 (1995).

\bibitem{Kis96} Z. Kis, J. Janszky, P.Adam, A.V. Vinogradov, and T. Kobayashi,
``Entangled vibrational states in polyatomic molecules,'' Phys. Rev. A 
{\bf 54}, 5110 (1996). 

\bibitem{Jan94} J. Janszky, A.V. Vinogradov, T. Kobayashi, and Z. Kis,
``Vibrational Schr\"odinger-cat states,'' Phys. Rev. A {\bf 50}, 1777 (1994).

\end{thebibliography}
\end{document}